\RequirePackage{fix-cm}
\documentclass[twocolumn,epjc3]{svjour3}  
\smartqed  
\RequirePackage{graphicx}
\RequirePackage{amsmath}
\RequirePackage{caption}
\RequirePackage{subcaption}
\RequirePackage{enumitem}
\RequirePackage[english]{babel}
\RequirePackage[autostyle, english = american]{csquotes}
\RequirePackage[switch]{lineno}

\MakeOuterQuote{"}

\RequirePackage[colorlinks,citecolor=blue,urlcolor=blue,linkcolor=blue]{hyperref}

\journalname{Eur. Phys. J. C}

\begin{document}
\title{Reconstruction of cosmic-ray muon events with CUORE}
 
\author{ D.~Q.~Adams\thanksref{USC} \and C.~Alduino\thanksref{USC} \and K.~Alfonso\thanksref{LBNLNucSci,VirginiaTech} \and A.~Armatol\thanksref{LBNLNucSci} \and F.~T.~Avignone~III\thanksref{USC} \and O.~Azzolini\thanksref{INFNLegnaro} \and G.~Bari\thanksref{INFNBologna} \and F.~Bellini\thanksref{Roma,INFNRoma} \and G.~Benato\thanksref{GSSI,LNGS} \and M.~Beretta\thanksref{Milano,INFNMiB}
% ...
} 

\institute{ Department of Physics and Astronomy, University of South Carolina, Columbia, SC 29208, USA\label{USC} \and Nuclear Science Division, Lawrence Berkeley National Laboratory, Berkeley, CA 94720, USA\label{LBNLNucSci} \and Center for Neutrino Physics, Virginia Polytechnic Institute and State University, Blacksburg, Virginia 24061, USA\label{VirginiaTech} \and INFN -- Laboratori Nazionali di Legnaro, Legnaro (Padova) I-35020, Italy\label{INFNLegnaro} \and INFN -- Sezione di Bologna, Bologna I-40127, Italy\label{INFNBologna} \and Dipartimento di Fisica, Sapienza Universit\`{a} di Roma, Roma I-00185, Italy\label{Roma} \and INFN -- Sezione di Roma, Roma I-00185, Italy\label{INFNRoma} \and Gran Sasso Science Institute, L'Aquila I-67100, Italy\label{GSSI} \and INFN -- Laboratori Nazionali del Gran Sasso, Assergi (L'Aquila) I-67100, Italy\label{LNGS} \and Dipartimento di Fisica, Universit\`{a} di Milano-Bicocca, Milano I-20126, Italy\label{Milano} \and INFN -- Sezione di Milano Bicocca, Milano I-20126, Italy\label{INFNMiB}
% ...
}

\thankstext{fn1}{Deceased} \thankstext{fn2}{Deceased} \thankstext{fn3}{Presently at: Instituto de Física, Universidade de São Paulo, São Paulo 05508-090, Brazil}

\date{Received: date / Accepted: date}

\maketitle

\begin{abstract}
We report the in-situ 3D reconstruction of through-going muons in the CUORE experiment, a cryogenic calorimeter array searching for neutrinoless double beta ($0\nu\beta\beta$) decay, leveraging the segmentation of the detector. Due to the slow time response of the detector, time-of-flight estimation is not feasible. Therefore, the track reconstruction is performed using a multi-objective optimization algorithm that relies on geometrical information from the detector as a whole. We measure the integral flux of cosmic-ray muons underground at the {\it Laboratori Nazionali del Gran Sasso}, and find our value to be in good agreement with other experiments that have performed a similar measurement. To our knowledge, this work represents the first demonstration of 3D particle tracking and reconstruction of through-going muons with per-event angular determination in a millikelvin cryogenic detector array. The analysis performed for this work will be critical for validating the muon-related background in CUPID, a next-generation $0\nu\beta\beta$ experiment, and for follow-up studies on detector response and on delayed products induced by cosmic-ray muons.

\end{abstract}

\section{Introduction}
\label{intro}
High-energy cosmic rays interacting with the Earth's atmosphere produce secondary particles propagating towards the Earth's surface. Among these particles are pions and kaons that eventually decay to produce cosmic-ray muons. These muons will reach the surface of the Earth and, if energetic enough, reach underground sites where physics experiments are located. These underground experiments often search for rare events that require an ultra-low background level, and cosmic-ray can constitute a significant source of background events to some searches. Quantifying this background is critical to reach physics goals in underground searches for rare and novel particles and interactions.

\begin{figure}[htp!]
  \includegraphics[width=\columnwidth]{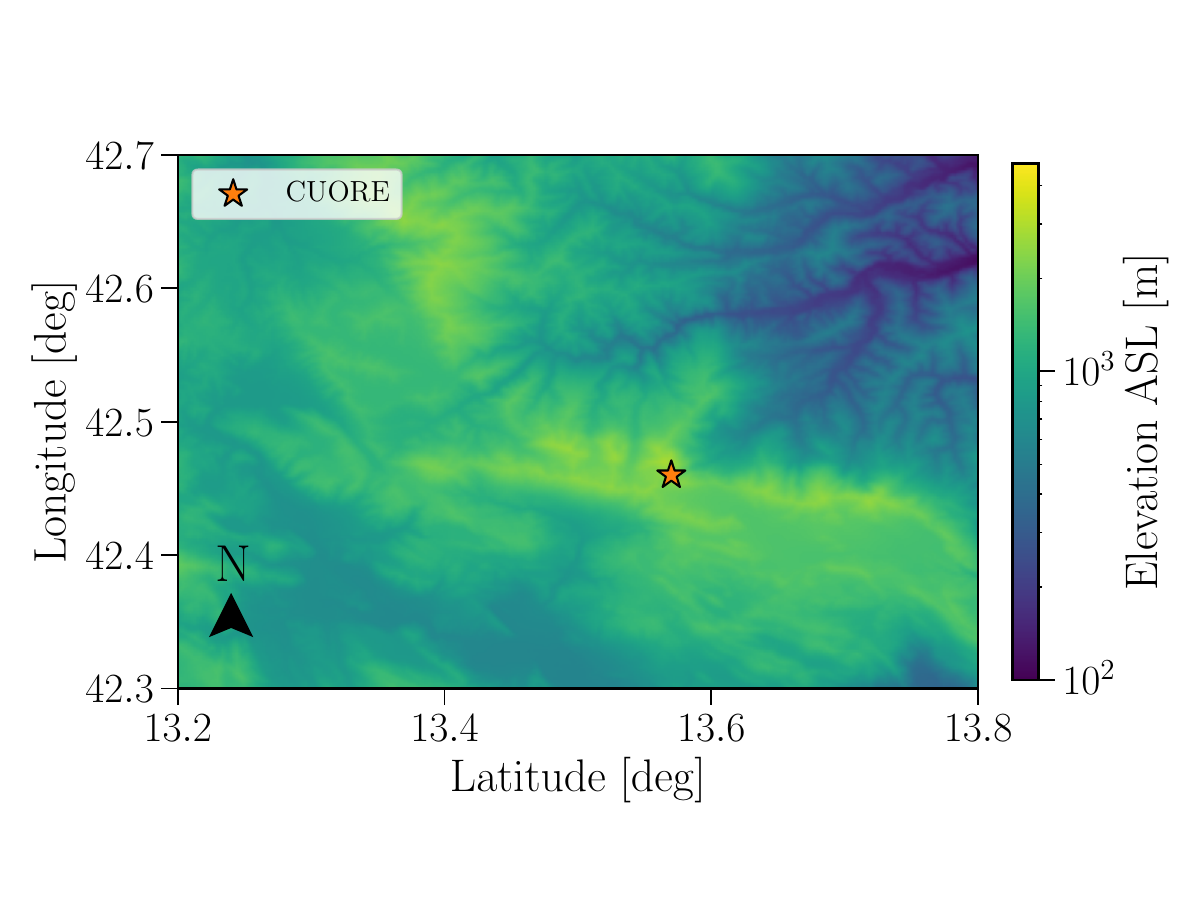}
\caption{Topographical map of the area surrounding LNGS where the CUORE experiment is located, displaying elevation Above Sea Level (ASL). Data obtained from the United States Geological Survey Website~\cite{US_GeologicalSurvey}.}
\label{fig:topoMap}       
\end{figure}

One of these underground experimental sites is the {\it Laboratori Nazionali del Gran Sasso} (LNGS), located under the Gran Sasso mountains in central Italy. The mountain profile over the site---shown in Fig.~\ref{fig:topoMap}---provides an asymmetric overburden of about 3600 meter-water-equivalent (m.w.e.), reducing the underground muon flux by roughly six orders of magnitude compared to the surface~\cite{Borexino:2012wej,MACRO}. 

The Cryogenic Underground Observatory for Rare Events (CUORE) is a cryogenic-calorimeter array located at Hall A of LNGS, primarily searching for the neutrinoless double beta decay of $^{130}$Te. While optimized towards the search for $0\nu\beta\beta$ decay in $^{130}$Te, CUORE has reached sufficient scale to reconstruct track-like events such as muons crossing the detector's volume. This capability opens the door for new physics avenues with CUORE, such as searches for track-like exotic phenomena, as demonstrated in~\cite{CUORE:2024rbd}. CUORE observes muon-induced events at a rate of a few per hour. Typical signatures are throughgoing tracks that may be accompanied by showers of secondary radiation caused by the interaction of the primary muon with detector components.

Several measurements of the underground muon flux at LNGS have been performed with experiments optimized for high geometric acceptance and large volumes. These experiments are typically instrumented with photodetectors with excellent timing resolution, and can perform time-of-flight measurements of a through-going particle and aid in track reconstruction. This, in turn, improves data selection so they can achieve high purity in their sample. In contrast, CUORE's size is significantly smaller than other underground experiments targeting astroparticle studies, yielding a lower statistical sample than those of previous measurements. Nonetheless, CUORE's segmented design and sensitivity to energy depositions at $\mathcal{O}(10\,\text{keV})$ thresholds has enabled searches for beyond-the-Standard Model track-like signatures~\cite{CUORE:2024rbd}. CUORE's site and cryogenic infrastructure will host a next-generation $0\nu\beta\beta$ experiment: the CUORE Upgrade with Particle Identification (CUPID)~\cite{CUPID:2025avs}. Characterizing the muon background at the CUORE site and validating related simulation and analysis tools will be of utmost importance to allow CUPID to reach its science goals.

In this analysis, we use through-going muons to perform an in situ muon-flux measurement using the CUORE detector. We focus on preparing a track-exclusive sample of muons that pass directly through the detector's channels to resolve the differential flux of muons underground at LNGS and measure their integral flux. We take advantage of the experiment's segmentation for the event selection, keeping only high-crystal-multiplicity and high-energy events. These events will appear as tracks, which are then fit by a multi-optimization algorithm from which we infer the candidate-muon direction.

\section{The CUORE experiment}

CUORE is an ultra-low background~\cite{CUORE:2024fak} experiment located in Hall A of LNGS primarily searching for neutrinoless double beta decay ($0\nu\beta\beta$). It consists of 988 TeO$_2$ crystals arranged into 19 towers, each with 4 columns of 13 crystals, and covering a volume of $\sim$1\,m$^3$ as represented in Fig.~\ref{fig:mayavi_rendering}. CUORE's crystals serve as both a decay source and a detection medium for the search of $0\nu\beta\beta$ in $^{130}$Te. The crystals are instrumented with neutron transmutation-doped (NTD) Germanium thermistors to register temperature changes in the crystal absorber upon energy deposition by an incident particle. These detectors require cryogenic temperatures to operate properly and are kept at $\sim$10\,mK base temperature by a large dilution refrigerator~\cite{CuoreCryo}. This thermal readout technique provides strong sensitivity to energy deposition within the crystal absorbers, regardless of particle type. CUORE's segmentation can select signatures triggering many calorimeters in coincidence with each other, or be used to reduce background to single-crystal events through an anticoincidence veto.

CUORE began taking data in April 2017, with data collection continuing to the present. CUORE's latest data release analyzed over 2 tonne-years of TeO$_2$ exposure~\cite{CUORE:2024ikf}, and currently holds the best limit on the half-life of $0\nu\beta\beta$ decay in $^{130}$Te~\cite{CUORE:1TY}. The data are divided into intervals of about 1-2 months of continuous background data collection, which we refer to as ``datasets'' for the rest of the text. Datasets consist of data-taking subsets referred to as ``runs,'' roughly of 1-day duration. Each dataset is preceded and followed by a calibration period of about one week, in which radioactive sources are lowered around the CUORE cryostat to obtain calibration data and subsequently characterize the detector response. The 2 tonne-year data release comprised 28 datasets numbered \texttt{1}--\texttt{28} and collected between May 2017 and April 2023. For this work, we use background runs from the 2 tonne-year data release, excluding datasets \texttt{1} and \texttt{5} due to insufficient detector-wide performance under the selections used for this analysis, corresponding to a cumulative 2.73 years of detector live-time.

Data processing within CUORE begins with each crystal acting as an independent detector, and continuously collecting waveforms in a trigger-less manner for subsequent offline analysis. A new addition to the data pipeline for the 2 tonne-year data release is a denoising algorithm~\cite{Vetter:2023fas,adams2025endtoenddataanalysismethods}. This algorithm uses information collected by auxiliary acoustic and mechanical devices such as microphones, seismometers, and accelerometers to subtract correlated noise across the detector. Subsequently, these denoised waveforms are then retriggered offline with an algorithm based on an optimum filter~\cite{OptimumFilter} designed to maximize the signal-to-noise ratio and decrease the energy threshold. The pulse amplitude is then evaluated on this filtered waveform and subsequently stabilized by correcting for gain fluctuations caused by thermal drifts of the detector~\cite{CUORE:analysisTech}. The next processing step converts the stabilized amplitude to units of energy by utilizing the data from calibration runs, in which physics peaks of known energy are resolved. Specific to this analysis, we additionally include in our selections pulses from events that saturate detector electronics. These events are typically discarded as current processing algorithms used by CUORE cannot determine their energy due to their truncated pulse shapes, although time-above-threshold methods are being considered to recover their energies in future work~\cite{adams2025endtoenddataanalysismethods}. Nonetheless, these saturated pulses still provide important spatial information to reconstruct high-energy muon tracks.

\section{Track reconstruction with Multiple Object Optimization}
\label{sec:trackFit}
For this work, we use the multi-objective optimization (MOO) algorithm developed in~\cite{Yocum:2022hum} to assign a best-fit track path to a collection of coincident events within the detector, implemented using the \texttt{Python} library \texttt{pymoo}~\cite{pymoo}. The algorithm was developed for CUORE to use a holistic approach that uses information from the complete detector array, including the position and energy information of all channels, even those that do not present energy deposition, to reconstruct a track from a muon candidate. The advantage of this algorithm over a simple approach, such as a least-squares reconstruction, is that it is more robust against track-fitting biases due to accidental coincidences or stray radiation that would pull the reconstructed track away from the true track.

\begin{figure}
  \includegraphics[width=\columnwidth]{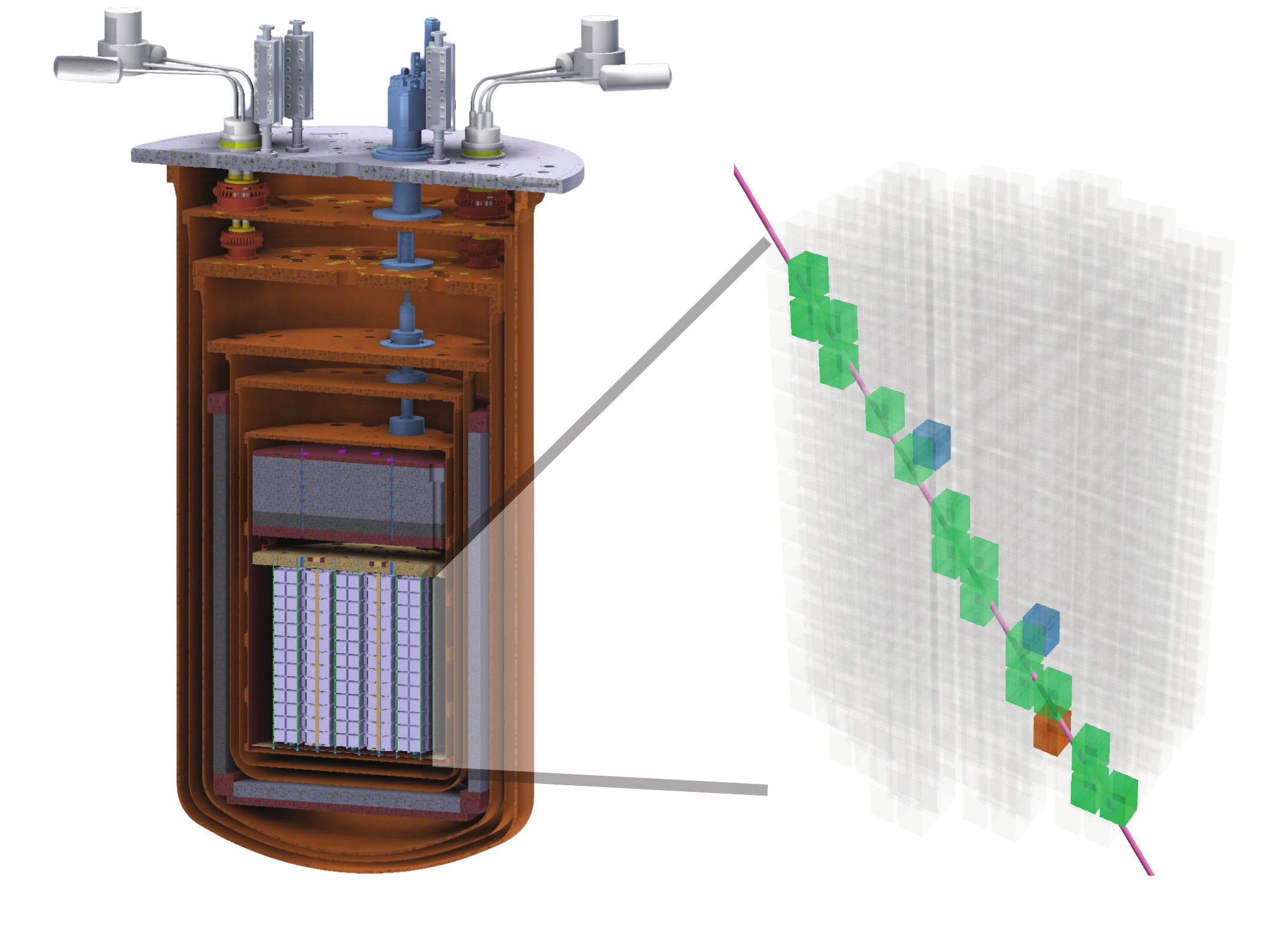}
\caption{An observed muon candidate through CUORE, after event selection and fitting have been applied to data. The purple line represents the best track obtained with the multi-objective optimization algorithm described in Sec.~\ref{sec:trackFit}. The green cubes represent hit channels, blue cubes are missed channels, and red cubes are extra channels within the data as described in Sec.~\ref{sec:trackFit}. Rendering produced with \texttt{mayavi}~\cite{Ramachandran2011}.}
\label{fig:mayavi_rendering}       
\end{figure}

We group contemporaneous energy depositions across multiple detector channels (either simulated or within our detector data) into a single detector-wide event, called a cluster. We set an analysis energy threshold of $9\,$MeV, as described in Sec~\ref{subsec:event_selection}. It is useful to define the following quantities extracted by the MOO track-reconstruction algorithm for a particular cluster, defined relative to the best-fit track:
\begin{itemize}
    \item "Hit channels": detectors that register energy deposits greater than the analysis threshold and are intersected by the best-fit track.
    \item "Missing channels": detectors missed by the reconstructed track that registered energy deposits greater than the analysis threshold.
    \item "Extra channels": detectors intersected by the reconstructed track but did not register energy deposition above the analysis threshold.
\end{itemize}
Examples of these quantities are shown in Fig.~\ref{fig:mayavi_rendering} for a sample muon candidate within CUORE. During fit evolution, the MOO track-reconstruction algorithm minimizes differentiable approximations to the number of missing and extra channels, as these two classes of channels represent departures from a perfect track topology. For muon events passing through the detector, some missing and extra channels occur due to secondary particles and detector effects. For example, extra channels can be attributed to crystals that were excluded from the analysis after selection cuts due to energies lower than the analysis threshold for this work (e.g., muons depositing energy below our analysis threshold as they ``clip'' the corner of a crystal) or exclusion due to suboptimal performance of a channel during that period. Missing channels occur when radiation carried away from the primary particle deposits energy above the analysis threshold. These three quantities are used to validate the reconstruction algorithm by comparing data and Monte Carlo simulations. Other quantities extracted from this algorithm are the azimuth and zenith angle of the through-going muon candidate, where both angles are determined assuming the track to be down-going.

Only clusters remaining after event selection (c.f. Sec.~\ref{sec:event_selection}) are track-fitted, as the algorithm has a considerable computational overhead. We benchmark the performance and angular reconstruction resolution of the reconstruction algorithm using simulated events with CUORE's \texttt{Geant4}-based~\cite{Geant4,Geant4_2} simulation package. These muon events follow the distribution measured by the MACRO experiment, and are simulated on a 5\,m-radius hemisphere centered around the cryogenic-detector array. They are subsequently propagated through CUORE and its external shielding using a realistic detector geometry, recording the simulated energy deposition within CUORE crystals.

\begin{figure}[htp!]
  \includegraphics[width=\columnwidth]{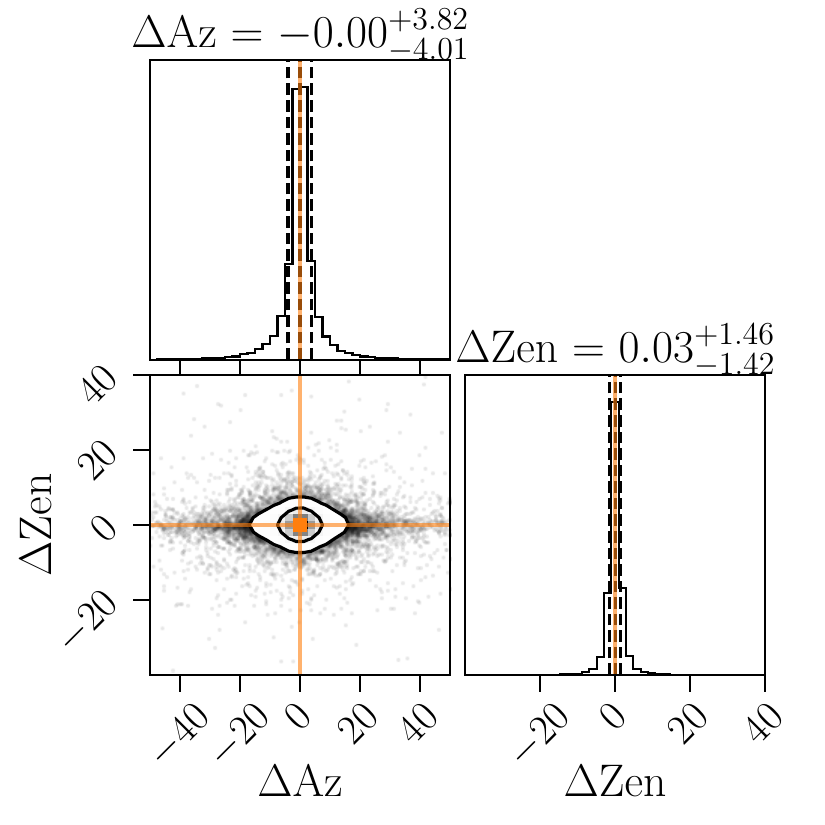}
\caption{Angular reconstruction accuracy (in degrees) and $1\sigma$ levels of the reconstruction algorithm obtained with MC simulations of muons in the CUORE experiment. $\mathbf{\Delta}$ denotes (\texttt{true}-\texttt{reconstructed}), where \texttt{true} refers to values obtained from our Monte Carlo simulations, whereas \texttt{reconstructed} refers to values inferred by the algorithm. Plot made with \texttt{corner}~\cite{corner}.}
\label{fig:cornerPlot_MCBenchmark}       
\end{figure}

We compare the reconstructed muon direction to the true value recorded from Monte Carlo simulations. We find that the pointing mean error and spread ($1\sigma$ level) in degrees are $\Delta\text{Az} = 0.00^{+3.82}_{-4.01}$ degrees and $\Delta\text{Zen} = 0.03^{+1.46}_{-1.42}$ degrees, as shown in Fig.~\ref{fig:cornerPlot_MCBenchmark}. The angular resolution differences between azimuth and zenith angles are due to the different lengths inter-crystal separation (for the zenith case) and to tower separation (for the azimuthal angle), the latter being greater than the former. This limited resolution is expected from CUORE's small footprint. The non-gaussian tails come from events with small zenith angles that might only trigger crystals in a single tower, and thus the algorithm would misreconstruct the azimuth angle. 

\section{Event selection}
\label{sec:event_selection}

Muon events that pass through the CUORE detector deposit energy on the crystals they intersect via continuous energy loss mechanisms, largely ionization. They also produce electromagnetic showers, creating secondary particles that reach crystals away from the primary track, distorting the track-like signature we wish to reconstruct. These secondaries, usually yield lower energy depositions in a crystal, and the original track can be reconstructed with good fidelity by selecting only the highest energy depositions. This criterion is applied when choosing the cuts in the event selection procedure.

\subsection{Event selection procedure}
\label{subsec:event_selection}

\begin{figure*}[htbp]
\centering
\begin{subfigure}{.45\textwidth}
  \centering
  \includegraphics[width=.8\linewidth]{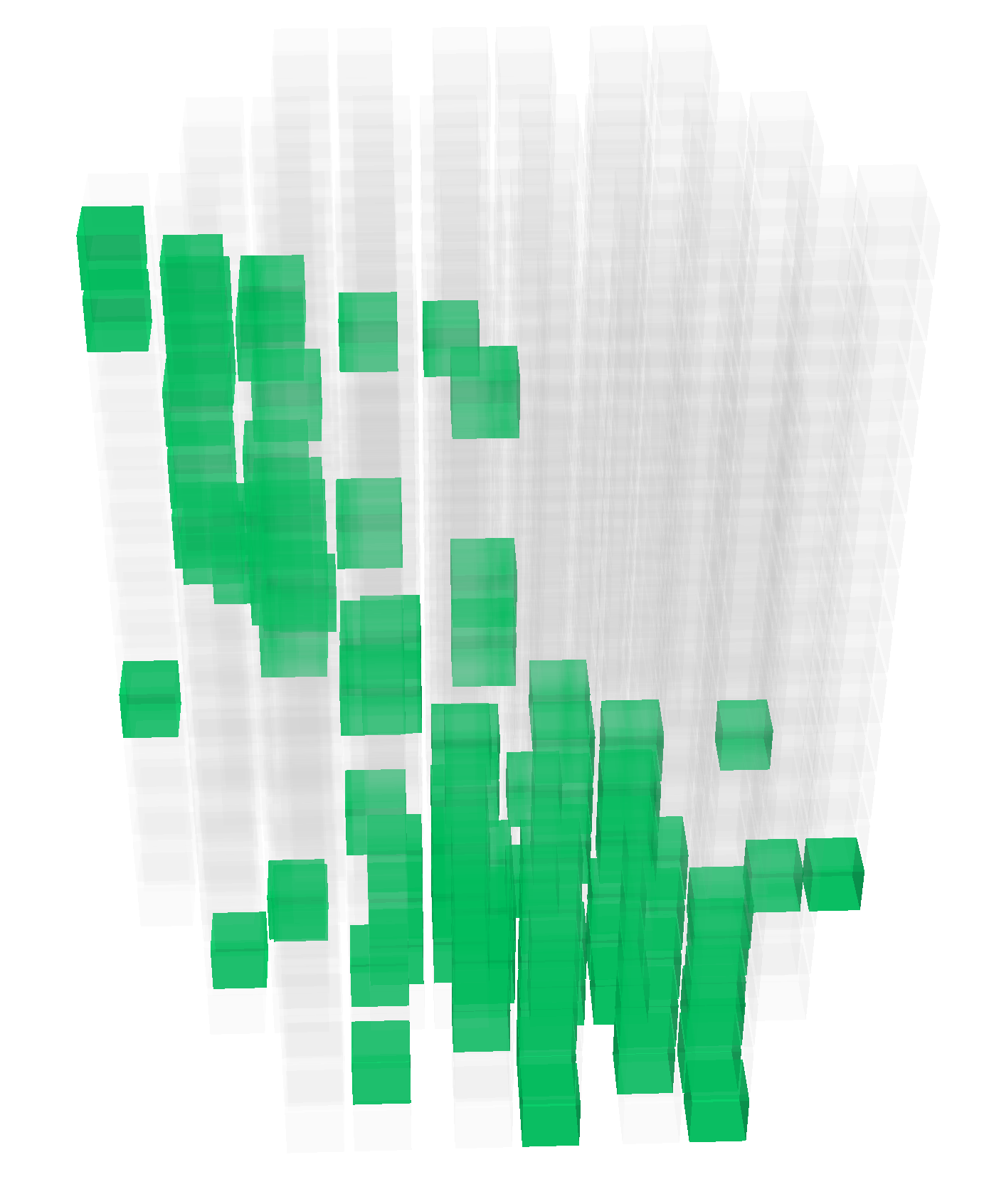}

  \label{fig:NoEnergyCut}
\end{subfigure}%
\hfill
\begin{subfigure}{.45\textwidth}
  \centering
  \includegraphics[width=.8\linewidth]{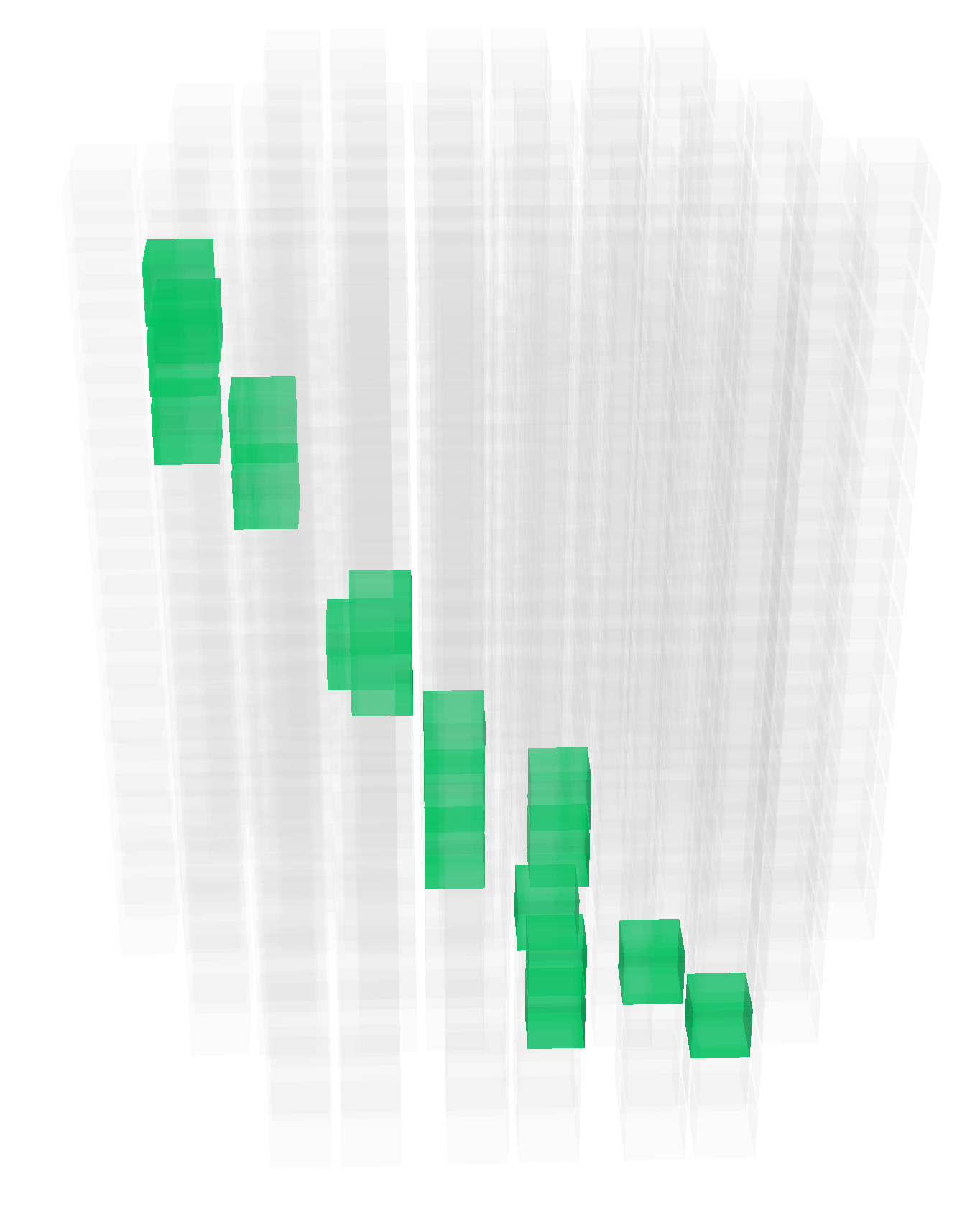}

  \label{fig:9MeVCut}
\end{subfigure}
\caption{Event topologies of a muon candidate in CUORE displayed with a 40\,keV analysis threshold (\textbf{left}) on triggered crystals (green) versus a 9\,MeV threshold analysis threshold (\textbf{right}). The track-like nature of the event is enhanced after discarding crystals that register lower-energy depositions due to secondary particles produced along the track.}
\label{fig:evTopologies}
\end{figure*}

We first apply basic cuts to remove spurious events, unphysical events, calibration pulses, and events that occurred during intervals when the detector was not taking data optimally (for example, due to earthquakes or maintenance activity underground near the CUORE experiment). We follow the cuts applied in CUORE's latest $0\nu\beta\beta$ decay analysis~\cite{CUORE:2024ikf}, excluding the Principal Component Analysis (PCA) cut, which rejects events based on the pulse-shape deviation from a channel-dependent average pulse calculated for each dataset. This PCA cut which would reject saturated muon events, which we do not wish to discard within this analysis.

In order to remove low energy secondaries and define an event selection primarily sensitive to throughgoing muon candidates, we set an energy analysis threshold of $>9$ MeV and also include saturated events with undetermined energy. This analysis threshold choice was selected considering the tradeoff between increasing the fraction of selected crystals which were directly intersected by a muon (as determined from Monte Carlo truth information), while maintaining good overall event statistics. An example of this can be seen in Fig.~\ref{fig:evTopologies}, where the left panel shows a topology of a muon candidate in data without cutting lower energy events, and the right panel shows the same event after channels below the analysis energy threshold are removed. It can be noticed that the removal of those low-energy events improves the track-like features of the event topology. We do not expect background events from any other standard model process to contribute events to our sample. All naturally occurring processes either occur at lower energies below these selections, or in the case of e.g. charged-current neutrino interactions within the detector itself, occur at negligible rates given the available exposure.

After these cuts, contemporaneous events that trigger different crystals are grouped into a detector-wide cluster. CUORE's thermal readout has intrinsically slow timing resolution, and event clustering is additionally complicated by timing offsets given by an inherent jitter distribution for each crystal. For this, a time-matched filter has been designed and used successfully in other high-crystal multiplicity analysis~\cite{CUORE:2024rbd}. This algorithm consists of a moving boxcar average that identifies periods consistent with multi-crystal events, grouping these events into a cluster, using a coincidence window of 100\,ms tuned to minimize the acceptance of false coincidences while maximizing the efficiency of true correlated events. We choose to keep only events whose multiplicity (that is, the number of distinct participating channels) is at least 5 ($\mathcal{M}\geq5$). This cut was selected to increase the purity of the muon sample and provide sufficient track lengths through the detector to provide reasonable pointing determinations.

Channel baselines in CUORE are monitored during data collection to maintain a dynamic range of up to $\sim 10$ MeV without saturating readout electronics. By extrapolating the amplitude of unsaturated pulses, we determine across the channel-exposures contributing to this analysis a median saturation threshold of $> 20$ MeV, with less than 0.2\% of channel-exposures registering as saturated at energies below our 9 MeV analysis threshold. We constrain possible miscalibration at our analysis threshold of 9 MeV using background peaks arising from the $\alpha$-decay of $^{212}$Po at 8.8 MeV (corresponding to the decay-$\alpha$ alone) and 8.9 MeV (also containing the corresponding nuclear recoil). We determine a bias shift of of $<0.4\%$ for the apparent mean value of these peaks relative to their nominal values, which is similar to previously measured behavior of TeO$_2$ cryogenic calorimeters to $\alpha$-decays~\cite{Bellini:2010iw}. We include as a systematic error the possibility that such a shift may affect the determination of our analysis threshold, which impacts our overall sample statistics of selected muon candidates by $~\sim 0.1\%$.

\subsection{Pre-analysis validation}
\label{sec:preAnalysis}
We initially select a small subset of our data for proof-of-concept and validation of the techniques and selections used in this analysis. To minimize the possibility of introducing any bias when tuning cuts, we select three datasets from different periods of CUORE's data-taking (corresponding to $\sim$10\% of the total exposure) that capture optimal and non-optimal data-taking situations. This is done to ensure that the cuts (c.f. Sec.~\ref{subsec:event_selection}) that we use to select our data are sufficient and adequate, and that the predicted behavior from simulation matches our observations for datasets that occur during selected data-taking periods.

\begin{figure}[htpb]
  \includegraphics[width=\columnwidth]{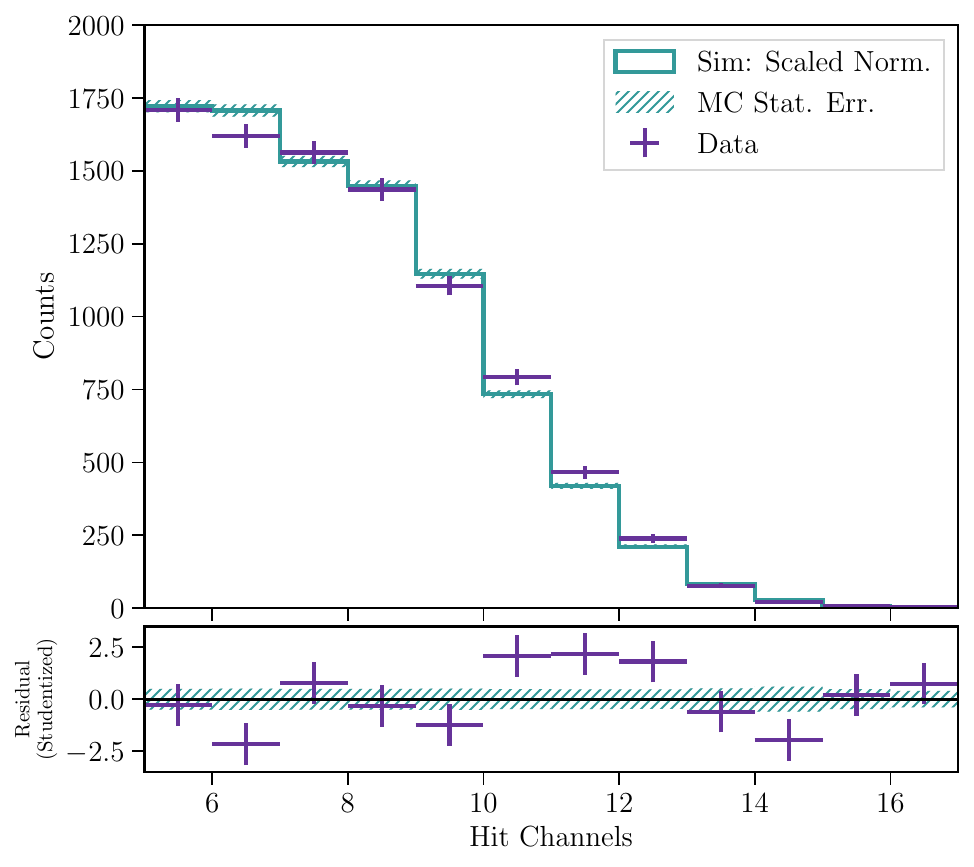}
\caption{Comparison between data and simulation of the hit channels distribution for muon candidates that pass our event selections. The simulation has been scaled to match the sample size and the joint 2-dimensional angular distribution as our observed data sample. The simulation and data distributions exhibit a $\chi^2$ comparison statistic of 19.0, for 11 degrees of freedom ($p=0.06$).}
\label{fig:HitChannelsComparison}       
\end{figure}

During this step, we tune our simulations to match the detector's channel-dependent deadtime for higher-multiplicity searches. This method is different than what is used for $0\nu\beta\beta$ searches, which focuses on single-site events, where the total detector exposure is given by the sum over per-channel exposures. For such low-multiplicity searches, when a particular channel is not active, it simply does not contribute to the total exposure. For searches focusing on events that trigger several channels, such as this analysis, one or more channels being offline for a subset of data-taking will not have a direct impact on the total detector exposure measured as a live-time, but will modify how an event will appear to the detector and the efficiency for detection. For example, the selection of a muon event which triggers a significant fraction of the channels in the detector will largely be unaffected by one or few channels being offline, but will have multiplicity observables modified by the channels that are offline when it occurred (Fig.~\ref{fig:HitChannelsComparison}).

Accordingly, we process our Monte Carlo simulations to include these effects. With our validation datasets, we scale our simulations according to the total detector livetime and the muon flux as previously measured by the Borexino experiment~\cite{Borexino:2012wej} to compare the marginal distributions of channel-multiplicity observables. We also compare the shape of the zenith and azimuthal angular distribution between our simulations and the data. This exercise showed discrepancies between the angular distributions for events reconstructed at small zenith angles (i.e., completely down-going muons). Upon investigation, these discrepancies were identified to be caused by spurious non-particle events. These spurious events suffer from unphysical waveforms that otherwise would have been identified by the PCA cut, which we do not use within this analysis; we do not expect these events to pass selections for other analyses which do employ this data-quality cut. The specific origin of these events remains unsolved, but current efforts are underway to understand and mitigate them. These events seem to occur in burst-like periods, leading to higher than nominal cluster rates within the detector when present. For this study, we assume a normally-distributed and stationary rate of true muon candidate events, and we choose to reject runs that had a rate greater than 21.3 clusters/run-day\footnote{Per-run rates were normalized to 24 hours, since runs do not have the same duration.} ($>2\sigma$), indicative of contamination by high rates of these spurious events. Such a rejection removes 8.3\% of the detector's pre-cut livetime from the analysis. This so-called $\mathcal{Z}$-cut has a small chance of rejecting an uncontaminated run, due to upwards statistical fluctuations in the per-run number of observed muons. We account for this bias by including the efficiency (and its associated $1\sigma$ uncertainty) to accept good runs as the ratio between the means of our determined good-run distribution, and that distribution truncated by this $\mathcal{Z}$-cut. This yields $\epsilon_{\mathcal{Z}}=0.955\, \, \pm \, \,0.014$, which enters into our measurement of the absolute flux.

\subsection{Impurity estimation}

 We inspect the inter-arrival time between identified muon candidate clusters, expecting the inter-arrival times of muons to be of the order of several minutes on average. Before removing the runs contaminated with spurious events described previously, we noticed a clear excess of clusters with short inter-arrival times. These correspond to rapid bursts of these events.

Before the analysis with our full selection statistics, we compute the possible residual impurity of the leftover sample after rejecting the events described in~\ref{sec:preAnalysis}. The residual impurity is estimated by fitting the inter-arrival time distribution of clusters for each dataset in two time windows, one in the region of small inter-arrival times (up to 2000 s) where we expect spurious events to be located, and a control region (greater than 2000 s) where most events correspond to muon candidate events. The fit of an exponential distribution to the control region is used to infer a muon-only rate, and is compared to the observed rate in the short bin, which may be contaminated by spurious events. Any deviation of the observed rate with respect to the predicted rate corresponds to the impurity (from spurious events) of a given dataset. We perform this estimation on a dataset-by-dataset basis, and choose to discard dataset \texttt{1}, as its purity was found to be less than $85\%$ with this method, resulting in a reduction of about 1\% in the muon sample statistics due to its small exposure. Across the rest of the datasets, we determine a purity fraction greater than $99\%$, with a combined impurity fraction compatible with zero. After this selection, we consider 996.3 days of detector livetime for this analysis.
\section{Analysis}
\label{sec:analysis}

\subsection{Clustering efficiency}

\begin{figure*}[htbp]
\centering
\begin{subfigure}{.4\textwidth}
  \centering
  \includegraphics[width=\linewidth]{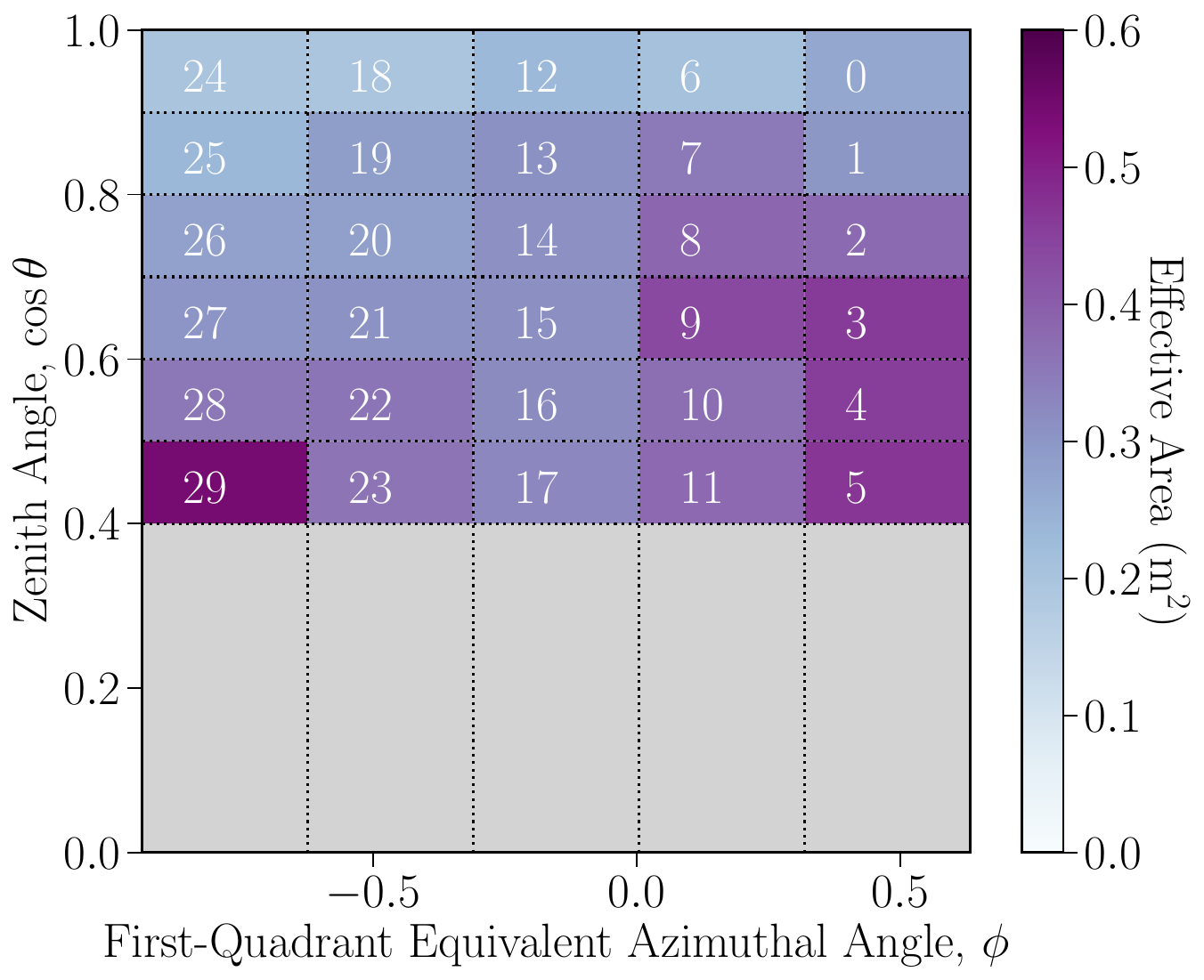}
  \label{fig:QuadAvg2DimEffectiveArea}
\end{subfigure}%
\hfill
\begin{subfigure}{.5\textwidth}
  \centering
  \includegraphics[width=\linewidth]{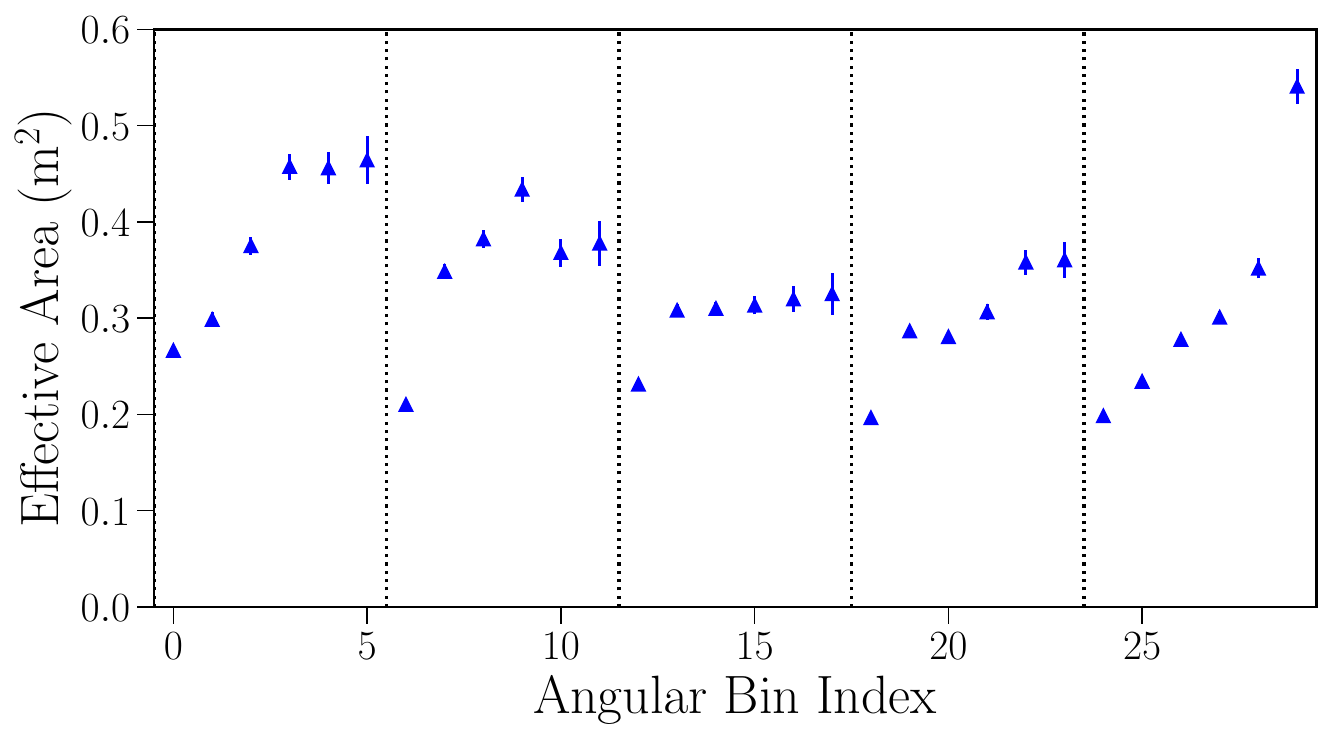}

  \label{fig:QuadAvg1DimEffectiveArea}
\end{subfigure}
\caption{The effective area of CUORE to muon events, including detector geometry and analysis-specific selections, as determined from Monte Carlo simulations. ({\bf Left}) Map of the differential effective detector area shown over two-dimensional angular bins in azimuthal and zenith angle. The azimuthal angle has been shown over the first quadrant of the detector, with the other quadrants related to the first quadrant by mirror symmetries in detector geometry, as shown in Fig.~\ref{fig:GeometryExplanation}. Zenith angles below $\cos \theta=0.4$ are not covered by the MC simulations, and are shown in gray. The corresponding index used to label each bin has been labeled within it in white. ({\bf Right}) The effective area of each angular bin, with 1$\sigma$ statistical uncertainty.  }
\label{fig:GeomEff}
\end{figure*}

As previously mentioned, clustering refers to grouping signals from coincidentally triggered crystals into a single detector-wide event of multiplicity $\mathcal{M}$. It uses a boxcar average with an optimized window of 100 ms. Using this algorithm, we wish to assess the probability that a complete cluster of $n$ triggered crystals will be grouped into an event of multiplicity $\mathcal{M}=n$. This is done by constructing mock clusters by resampling the distribution of trigger times occurring within $\pm100$ ms with respect to the mean trigger time of identified muon candidates. The width of the ordered times of the mock clusters is then compared against the 100 ms time window of the filter to estimate the efficiency. With this method, we estimate the clustering efficiency to be $\epsilon_{\text{clustering}}>0.995$ for $\mathcal{M}\geq5$.

\subsection{Effective Area and Geometric Efficiencies}
\begin{figure}[htpb]
  \includegraphics[width=\columnwidth]{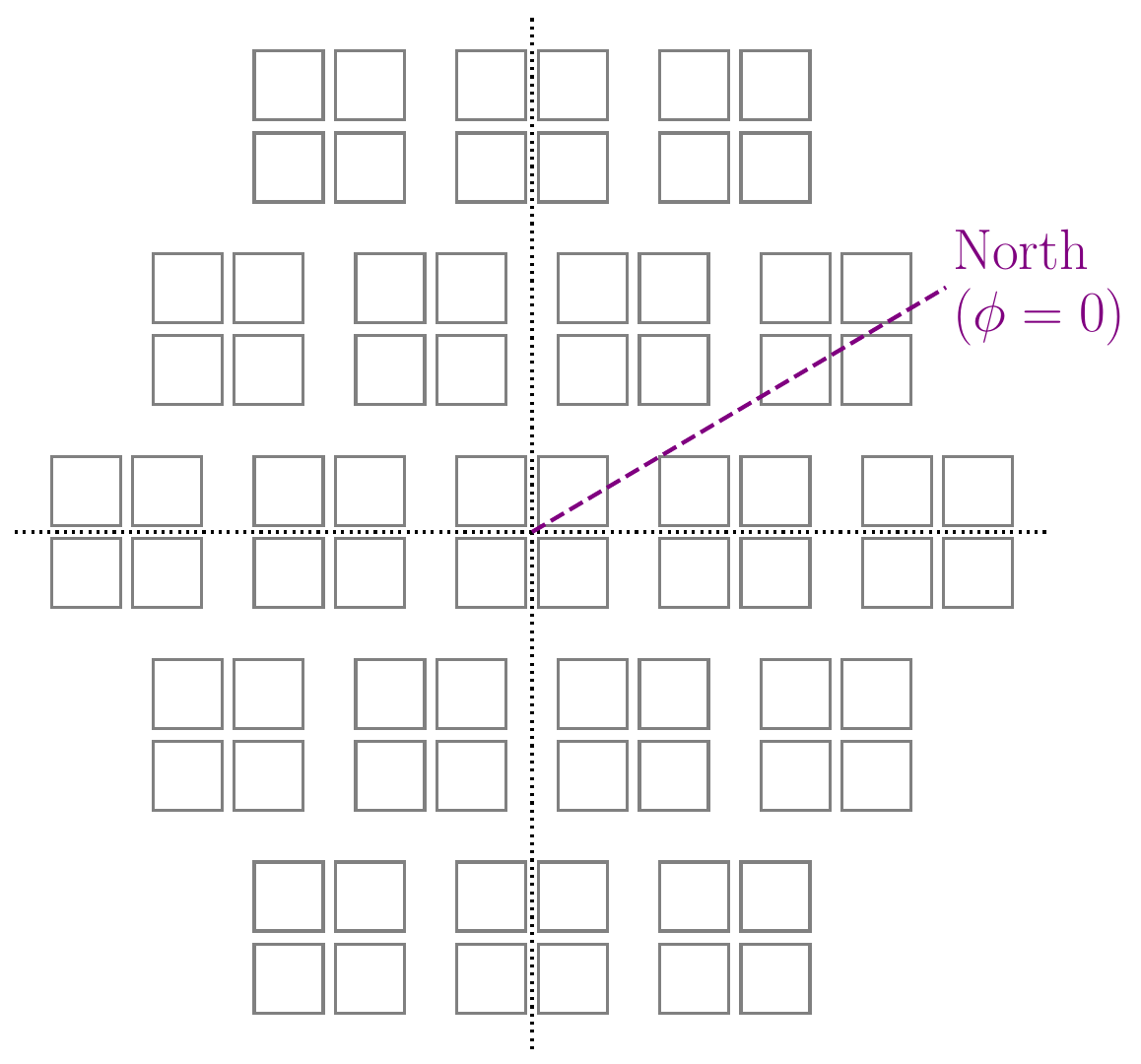}
\caption{Top-down layout of the CUORE detector tower geometry (gray boxes), relative to geographic North (purple). The azimuthal angle, $\phi$, is defined clockwise from geographic North. Geometric detector quadrants are separated by dashed lines, which correspond to the two mirror-planes of detector symmetry. The detector's construction is aligned with the principle directions of the LNGS experimental hall, offset by 36.24 degrees from geographic North~\cite{Andreino_Simonelli_2016}.}
\label{fig:GeometryExplanation}       
\end{figure}

CUORE's acceptance to cosmic ray muons varies with the spherical angle from which the detector is viewed, and depends on both the intrinsic detector geometry and the considered analysis selections.  We determine the effective area of the detector, $A_{\text{eff}}$, at a viewing angle of $(\phi,\cos \theta)$ from Monte Carlo simulations of the detector's response and geometry to cosmic-ray muons using CUORE's simulation package. Here, we use the convention that $\phi$ describes the azimuthal angle in radians from 0 to $2\pi$ defined clockwise from geographic North (as shown relative to the detector's layout in Fig.~\ref{fig:GeometryExplanation}), and that $\theta$ describes the zenith angle such that $\cos \theta = 1$ is vertically downwards and $\cos \theta =0 $ is horizontal. The MC simulated muon events are tuned on a dataset-dependent basis to reflect per-channel livetimes, followed by the same multiplicity selections as made to the detector data. The effective detector area we determine thus includes contributions from our analysis cut choices.

Practically, we define the effective area through geometric efficiencies, $\epsilon$, determined from our MC simulations:
\begin{equation}
    A_{\text{eff}} (\phi,\cos \theta) = \epsilon(\phi,\cos \theta) A_{\text{gen.}}.
\end{equation}
Here, $A_{\text{gen.}}$ is the total area of $67.5\, \, \pm \, \,0.8 $ m$^2$ covered by our cosmic ray muon generator, encompassing the entire footprint of the detector. We ascribe to the generator normalization a systematic uncertainty describing the slight non-homogeneity of the simulated flux over the vertical extent of the detector volume~\cite{Mayer:2024fvd}. We determine the geometric efficiencies per-dataset over angular bins in $\phi$ and $\cos\theta$ as
\begin{equation}
    \epsilon_\text{bin} = \frac{\#\{\text{Muons Passing Selections in bin\}}}{\#\{\text{Muons Simulated in bin\}}},
\end{equation}
The angular binning and overall effective area of CUORE are shown in Fig.~\ref{fig:GeomEff}. These binned values are then interpolated over $\phi$ and $\cos \theta$ with a linear spline over the centers of the angular bins. Due to a shortcoming in angular coverage within the used MC event generator, the simulations we use do not extend to zenith angles more shallow than $\cos \theta = 0.4$. We nominally consider a flat extrapolation of the detector's effective area from $\cos \theta = 0.4$ to $\cos \theta = 0$, and assess alternate extrapolation methods as a source of systematic error in Sec.~\ref{sec:syst}. Future MC data production campaigns will ensure full angular coverage of the detector.

We determine the overall effective area of our selections to be between 34-82\% that of an ideal cylinder (with $h\approx 0.37$ m, $r\approx 0.7$ m) circumscribed about the CUORE detector crystals, varying with viewing angle. The total flux-averaged effective area of our selections is found to be 0.339 m, 52\% of such a circumscribed cylinder. 

\subsection{Final sample}

\begin{figure*}[htbp]
\centering
\begin{subfigure}{.45\textwidth}
  \centering
  \includegraphics[width=\linewidth]{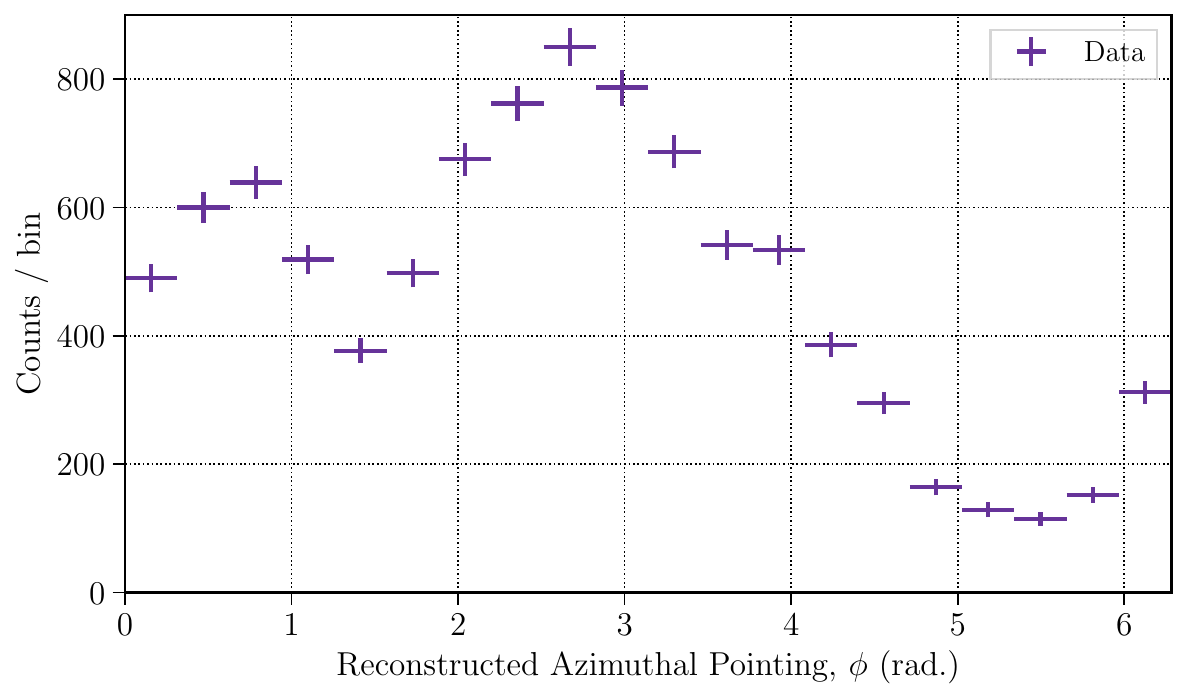}
  \label{fig:AzimuthPointing}
\end{subfigure}%
\hfill
\begin{subfigure}{.45\textwidth}
  \centering
  \includegraphics[width=.8\linewidth]{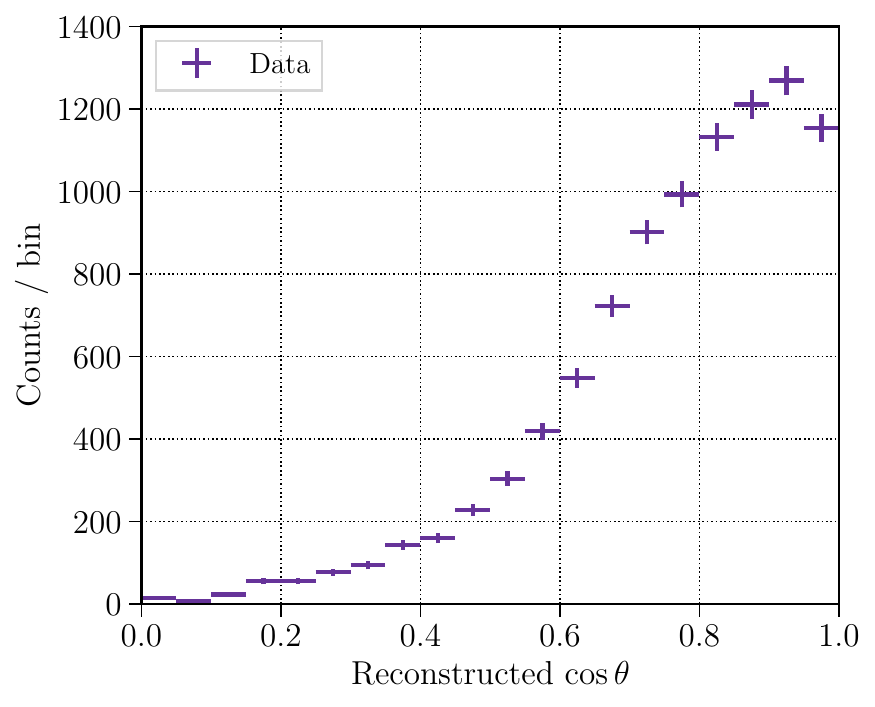}

  \label{fig:ZenithPointing}
\end{subfigure}
\caption{Distribution of the reconstructed azimuthal ({\bf left}) and zenith  ({\bf right}) direction of the final muon candidate sample. The local maxima and minima of the azimuthal angular distribution correlate with the topographical features of the site and asymmetrical overburden, as can be inferred from Fig.~\ref{fig:topoMap}. We note that the displayed azimuth phase and sign convention (clockwise, from geographic North) differs from that presented in \cite{PhysRevD.100.062002,Woodley:2024eln}, and that the observed distributions we present have not been corrected by angularly-dependent efficiencies.}
\label{fig:AngDistr}
\end{figure*}

\begin{figure}[htp!]
  \includegraphics[width=\columnwidth]{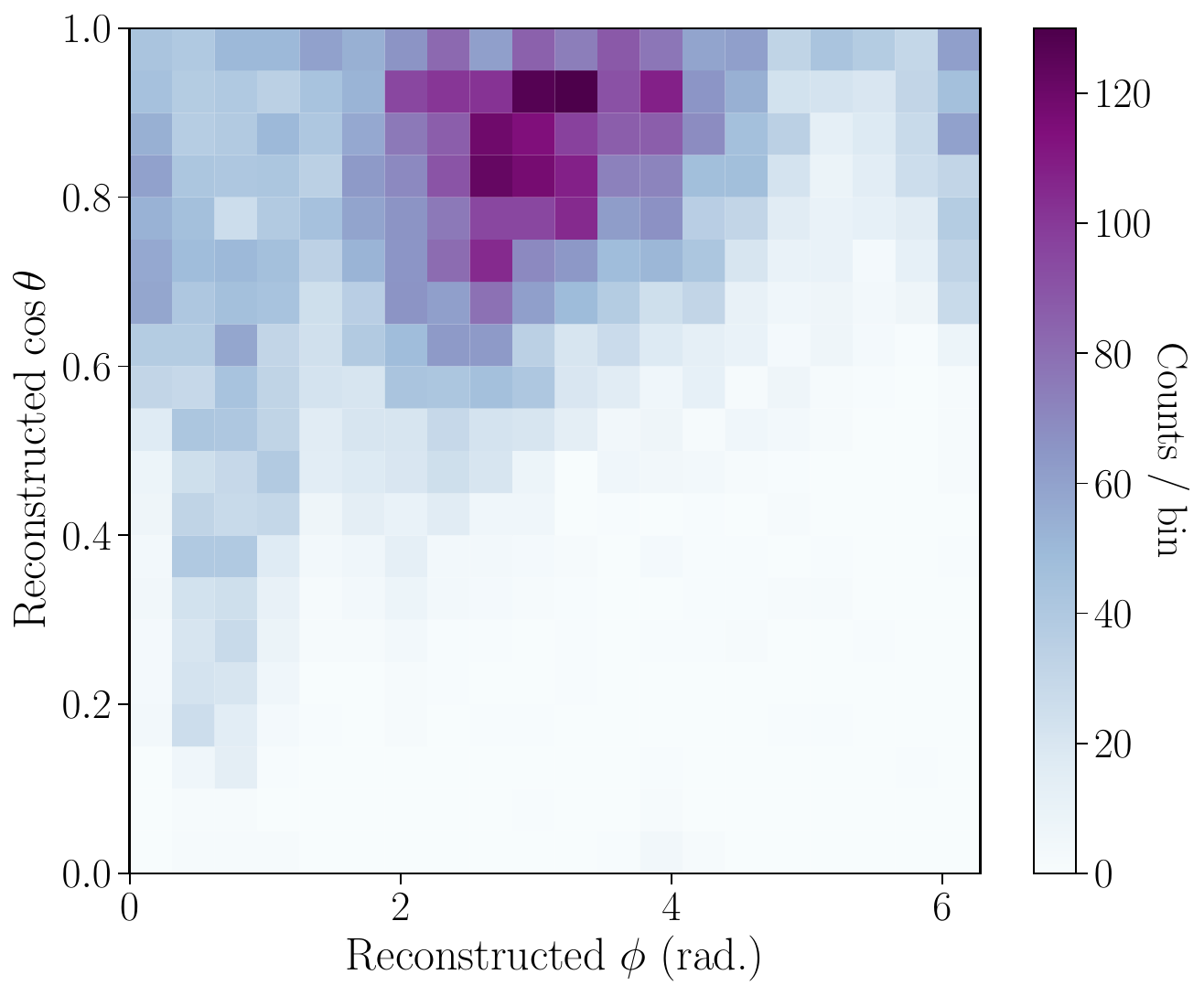}
\caption{Two-dimensional histogram of the reconstructed direction of all muon candidates within our sample. The angular variation, particularly with respect to the azimuthal $\phi$ angle, is due to the uneven overburden above LNGS.}
\label{fig:2DPointing}       
\end{figure}

The total number of events after all selection cuts have been applied and after problematic datasets and runs have been removed is 9513 muon candidates, which corresponds to $76.5\%$ of the initial sample. This sample is track-fitted using the algorithm described in Sec.~\ref{sec:trackFit}, from which variables such as the reconstructed direction for both azimuth and zenith angles are inferred. The unweighted angular distribution of muon candidates in the final sample is shown in Fig.~\ref{fig:AngDistr} and Fig.~\ref{fig:2DPointing}.

\section{Results}

\begin{figure*}[htbp]
\centering
    \includegraphics[width=\textwidth]{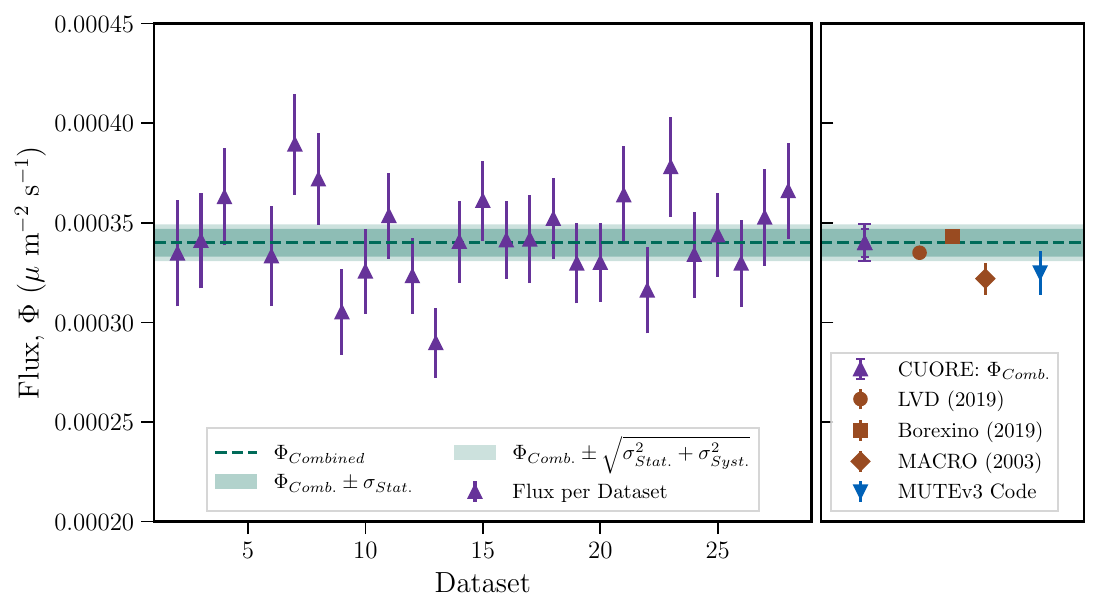}
    \caption{Measured flux as a function of the analyzed CUORE dataset. The green bands correspond to the uncertainty of the combined (dataset averaged) measurement, with different shades for only statistical uncertainty, statistical and systematic uncertainties added in quadrature and linearly. The right panel compares the flux obtained from this work to other experiments at LNGS (MACRO~\cite{MACRO}, LVD~\cite{PhysRevD.100.062002}, and Borexino~\cite{Borexino:2018pev}) and to the software package \texttt{MUTEv3}~\cite{Woodley:2024eln}.}
    \label{fig:FinalRateZoomed}
\end{figure*}

We first estimate the muon flux per dataset ($\Phi_{\text{ds}}$) as
\begin{equation}
\label{eq:rate}
    \Phi_{\text{ds}}= \frac{1}{\epsilon_{\mathcal{Z}} T}\frac{n}{\frac{1}{n}\sum_{i=1}^n \varepsilon_i A_\text{eff.}(\phi_i,\cos \theta_i)},
\end{equation}
where $T$ is the livetime of the dataset, $A_\text{eff.}(\phi_i,\cos \theta_i)$ 
is the effective area of CUORE to the $i^\text{th}$ muon candidate, and $\varepsilon_i$ its clustering efficiency as described in Sec.~\ref{sec:analysis}. The sum over $i$ runs over all observed muon candidates that pass all event selections, with the total number $n$. Note that these efficiencies are applied for each cluster in an unbinned sum. The flux per dataset (each of roughly 1.5 months duration) and its associated uncertainty can be seen on the left panel of Fig.~\ref{fig:FinalRateZoomed}. As shown in Table~\ref{tab:1} and Fig.~\ref{fig:FinalRateZoomed}, this analysis lacks the statistical precision to resolve the expected 1\%~\cite{Borexino:2018pev} seasonal modulation of the muon rate caused by atmospheric fluctuations across different datasets.

The overall muon flux is obtained by calculating the error-weighted average of the muon flux per dataset. This yields a measured flux of $\Phi=[3.40\, \, \pm \, \,0.07 \,(\text{stat.}) \, \, $ $\pm \, \,0.06\,(\text{syst})]\times10^{-4}$\,s$^{-1}$\,m$^{-2}$. Statistical sources of uncertainty are presented in Table~\ref{tab:1}, while the systematic contributions are listed in Table~\ref{tab:2} and discussed in Sec.~\ref{sec:syst}. We compare the flux calculated in this work with measurements from other experiments such as MACRO~\cite{MACRO}, Borexino~\cite{Borexino:2012wej}, and LVD~\cite{PhysRevD.100.062002}, along with their uncertainties. We also display the flux computed by the simulation package \texttt{MUTEv3}~\cite{Woodley:2024eln}. CUORE's small acceptance is unable to resolve spatially extended muon bundles (two or more muons passing through the laboratory simultaneously), whereas larger-footprint detectors can, which may lead to small differences in interpreting the overall rate, depending on whether the analysis is also inclusive to muon-bundles or exclusively single muons. Additionally, due to CUORE's comparatively poor time resolution ($\mathcal{O}$(ms)), time-of-flight determination is not feasible, making the detector unable to differentiate between down-going cosmic-ray muons vs. up-going muons from atmospheric neutrino interactions with the Earth's mantle. These events may account for $\mathcal{O}(1\%)$ of the observed rate~\cite{Bellini_2011}. The CUPID detector will be equipped with a muon veto~\cite{Moore:2025eil}, enabling discrimination between up- and down-going muons. Finally, smaller-effect discrepancies could be explained by the location of MACRO, BOREXINO, and LVD in different halls of LNGS, as well as by variation in the mean air temperature for the years from which the data was used for the measurements~\cite{Borexino:2012wej}.

\begin{table}
\centering
\caption{Statistics uncertainty sources and corresponding 1$\sigma$ impact on the estimated cosmic-ray muon flux. The total value that we quote was obtained by adding the uncertainties in quadrature.}
\label{tab:1}       
\begin{tabular}{lll}
\hline\noalign{\smallskip}
Source & Value ($10^{-6}$ m$^{-2}$ s$^{-1}$) & Rel. Value\\
\noalign{\smallskip}\hline\noalign{\smallskip}
$\mathcal{Z}$-cut &   5.0 & 1.5\% \\
Sample Statistics &  4.5 & 1.4\%\\
Clustering &   0.3 & 0.1\% \\
\noalign{\smallskip}\hline
Total & 6.8  & 2.0\% \\
\end{tabular}
\end{table}

\subsection{Systematic uncertainties}
\label{sec:syst}

We evaluate various sources of systematic uncertainty that may affect our flux measurement and efficiencies, which are summarized in Table~\ref{tab:2}. We determine these systematic errors by estimating the magnitude by which each considered source or effect impacts the final flux measurement, with the final reported systematic error taken to be the quadrature sum over the constituent errors, themselves conservatively symmetrized when necessary.

The dominant systematic error we determine originates from our inability to validate our analysis procedure at shallow zenith angles. As previously mentioned, our simulations exhibit a cut-off at $\cos \theta = 0.4$, representing a small ($<5\%$) but non-negligible portion of the muon flux. At a given azimuth angle, we nominally use a flat extrapolation of the effective area of CUORE from $\cos \theta =0.4$ to more shallow zenith angles. Considering only events with $\cos \theta > 0.4$ results in a sample of 9043 muon candidates, 4.9\% smaller than the total sample. As CUORE has a greater effective area at shallow angles, the partial flux resulting from this truncated sample is only 2.4\% smaller than our full measured rate. We alternatively consider possible different linear extrapolations to shallow angles as a systematic error.

For this, within each azimuth bin of Fig.~\ref{fig:GeomEff}, we take the effective area of the two shallowest zenith bins covered by the MC simulations: namely, over $\cos \theta \in [0.4,0.5]$ and $[0.5,0.6]$. We then re-throw the effective area for each of these last two bins by adding a Gaussian random variable with zero mean, and standard deviation equal to our 1$\sigma$ error determination of its value. These re-thrown values are then linearly extrapolated to obtain effective areas for $\cos \theta < 0.4$. To prevent unphysical extrapolations, we do not allow any extrapolated effective area to exceed that of a $h\approx 0.37$ m, $r\approx 0.7$ m cylinder which may circumscribe the array of CUORE crystals. To isolate effects coming from shallow-angle coverage, the rethrown values of the covered bins are only used for the purposes of extrapolation--the effective area of the last two covered bins are still taken as their centrally-determined values (the overall effect of perturbing the effective area of all bins is considered separately later in this section). We repeat this re-throw-and-extrapolate procedure 500 times, obtaining a sampling distribution of possible alternate, linear extrapolations of the effective area to shallow angles. The mean sample-averaged geometric efficiency arising from these repeated trials is found to be 1.4\% smaller than the nominally considered value, which correspondingly impacts our flux measurement through Eq.~\ref{eq:rate}.

Additionally, we find that different book-keeping methods lead to a small discrepancy in the total detector livetime considered for this analysis. This is due to each DAQ crate beginning and ending runs independently, leading to small ambiguities on the exact start and stop time for each crate with respect to the nominal run start/stop time. This effect is determined from the difference between the reported start and stop times from CUORE's database, as compared to Monte Carlo estimates of the livetime estimated by randomly querying whether detector channels were taking good data at a particular time. Summing over the 26 datasets we use for this analysis, the Monte Carlo estimates are found to report a value in total 0.6\% smaller than that queried from the database.

Further, we consider two different permutations of our analysis methodology to evaluate any possible impact on the final measured flux. First, we consider the difference between our nominal method of measuring the muon flux per-dataset and then determining a combined value, and the flux obtained from considering all muon events as if they originated from one effective dataset. This amounts to using Eq.~\ref{eq:rate}, where $T$ is the total considered livetime of the analysis, and the sum of $n$ runs over all muon events, using the original per-dataset estimates for the $\epsilon_i$ geometric efficiencies. This yields a flux value 0.5\% larger than our nominal quoted flux. Second, we evaluate the impact of using binned as opposed to interpolated values for the geometric efficiencies. This assigns to each muon candidate the mean value of the geometric efficiency of the angular bin it falls into, instead of linearly interpolating between bin centers. The resulting flux determined with binned geometric efficiencies is 0.1\% smaller than the quoted value using interpolated efficiencies.

\begin{table*}[htp]
\centering
\caption{Systematic uncertainty sources and impact on the estimated cosmic-ray muon flux. The total reported value is obtained by adding all the systematic uncertainties in quadrature.}
\label{tab:2}       
\begin{tabular}{lll}
\hline\noalign{\smallskip}
Source & Value ($10^{-6}$ m$^{-2}$ s$^{-1}$) & Rel. Value\\
\noalign{\smallskip}\hline\noalign{\smallskip}
Shallow Angle MC Coverage& 4.8 & 1.4\% \\
Generator Normalization& 3.2 & 0.9\%\\
Livetime& 1.9 & 0.6\% \\
Dataset Averaging& 1.6 & 0.5\% \\
Geom. Eff. Correlations & 0.6 & 0.2\% \\
Binning vs Interpolation & 0.3 & 0.1\% \\
Analysis Threshold & 0.3 & 0.1\% \\
\noalign{\smallskip}\hline
Total & 6.4 & 2.4\% \\
\end{tabular}
\end{table*}

Finally, we quantify the impact of correlated errors on the geometric efficiencies across different muon candidates. This stems from the fact that for two muon events occurring within the same dataset and with nearby azimuth and zenith values, any uncertainty in the geometric efficiencies for the two events will not be independent, but rather correlated. We estimate this effect by bootstrapping possible perturbations of the geometric efficiencies. For each bootstrapped trial, we adjust each geometric efficiency estimate (per dataset and angular bin) by adding a Gaussian random variable with zero mean, and standard deviation equal to our 1$\sigma$ error for that effective area. We then re-determine the overall flux. We consider the extra variance arising from these correlations by comparing to naive data resampling, which treats the error on the geometric efficiency for all muons as independent. We compare the two resulting distributions, and find that the distribution including correlations has a standard deviation 1.18 times larger than the distribution without, yielding a total relative increase in variance of 0.2\%.

\section{Summary}

The CUORE experiment, which targets $0\nu\beta\beta$ decay searches, has achieved analysis and hardware maturity for reconstructing track-like phenomena. In this work, we report a measurement on the flux of through-going muons at LNGS using high-multiplicity events from CUORE's latest two-tonne-year data release. We measure the total underground muon cosmic-ray flux at the CUORE site as $\Phi=[3.40\, \, \pm \, \,0.07 \,(\text{stat.}) \, \, \pm \, \,0.06\,(\text{syst})]\times10^{-4}$\,s$^{-1}$\,m$^{-2}$, which is consistent with the flux measured by other underground experiments at LNGS. Although this is not a precision measurement compared to that performed by experiments within LNGS having a larger effective area, this represents, to our knowledge, the first 3D reconstruction and cosmic-ray muon flux measurement by an array of cryogenic calorimeters. This also demonstrates the maturity and capabilities of cryogenic calorimeter arrays to perform searches and measurements of detector-wide astroparticle phenomena. Similarly, this analysis strengthens the confidence in the results of the latest CUORE search for fractionally charged particles using related methods~\cite{CUORE:2024rbd}. Finally, the muon sample obtained with this analysis will benefit follow-up background analyses such as searches for delayed cosmogenically activated events, detector response studies for high energy events, and for the study of secondaries for the validation of CUPID's background budget from muon-induced events.

\begin{acknowledgements}
The CUORE Collaboration thanks the directors and staff of the Laboratori Nazionali del Gran Sasso
and the technical staff of our laboratories.
This work was supported by the Istituto Nazionale di Fisica Nucleare (INFN);
the National Science Foundation under Grant Nos. NSF-PHY-0605119, NSF-PHY-0500337,
NSF-PHY-0855314, NSF-PHY-0902171, NSF-PHY-0969852, NSF-PHY-1307204, NSF-PHY-1314881, NSF-PHY-1401832, NSF-PHY-1913374, and NSF-PHY-2412377; Yale University, Johns Hopkins University, and University of Pittsburgh.
This material is also based upon work supported by the US Department of Energy (DOE)
Office of Science under Contract Nos. DE-AC02-05CH11231, and DE-AC52-07NA27344;
by the DOE Office of Science, Office of Nuclear Physics under Contract Nos.
DE-FG02-08ER41551, DE-FG03-00ER41138, DE-SC0012654,\\ DE-SC0020423, DE-SC0019316, and DE-SC0011091.
This research used resources of the National Energy Research Scientific Computing Center (NERSC).
This work makes use of both the DIANA data analysis and APOLLO data acquisition software packages,
which were developed by the CUORICINO, CUORE, LUCIFER, and CUPID-0 Collaborations.
The authors acknowledge the Advanced Research Computing at Virginia Tech and the Yale Center for Research Computing for providing computational resources and technical support that have contributed to the results reported within this paper.
\end{acknowledgements}

\bibliographystyle{spphys}       
\bibliography{refs}   

\begin{thebibliography}{10}
\providecommand{\url}[1]{{#1}}
\providecommand{\urlprefix}{URL }
\expandafter\ifx\csname urlstyle\endcsname\relax
  \providecommand{\doi}[1]{DOI \discretionary{}{}{}#1}\else
  \providecommand{\doi}{DOI \discretionary{}{}{}\begingroup \urlstyle{rm}\Url}\fi

\bibitem{Borexino:2012wej}
G.~Bellini, et~al., JCAP \textbf{05}, 015 (2012).
\newblock \doi{10.1088/1475-7516/2012/05/015}.
\newblock \urlprefix\url{https://doi.org/10.1088/1475-7516/2012/05/015}

\bibitem{MACRO}
M.~Ambrosio, et~al., Phys. Rev. D \textbf{56}, 1418 (1997).
\newblock \doi{10.1103/PhysRevD.56.1418}.
\newblock \urlprefix\url{https://link.aps.org/doi/10.1103/PhysRevD.56.1418}

\bibitem{CUORE:2024rbd}
D.Q. Adams, et~al., Phys. Rev. Lett. \textbf{133}(24), 241801 (2024).
\newblock \doi{10.1103/PhysRevLett.133.241801}.
\newblock \urlprefix\url{https://link.aps.org/doi/10.1103/PhysRevLett.133.241801}

\bibitem{CUPID:2025avs}
K.~Alfonso, et~al., Eur. Phys. J. C \textbf{85}(7), 737 (2025).
\newblock \doi{10.1140/epjc/s10052-025-14352-1}.
\newblock \urlprefix\url{https://doi.org/10.1140/epjc/s10052-025-14352-1}

\bibitem{US_GeologicalSurvey}
{U.S. Geological Survey}.
\newblock Global multi-resolution terrain elevation data 2010 ({GMTED2010}).
\newblock \urlprefix\url{https://topotools.cr.usgs.gov/gmted_viewer/index.html}

\bibitem{CUORE:2024fak}
D.Q. Adams, et~al., Phys. Rev. D \textbf{110}(5), 052003 (2024).
\newblock \doi{10.1103/PhysRevD.110.052003}.
\newblock \urlprefix\url{https://doi.org/10.1103/PhysRevD.110.052003}

\bibitem{CuoreCryo}
C.~Alduino, et~al., Cryogenics \textbf{102}, 9 (2019).
\newblock \doi{10.1016/j.cryogenics.2019.06.011}.
\newblock \urlprefix\url{https://doi.org/10.1016/j.cryogenics.2019.06.011}

\bibitem{CUORE:2024ikf}
D.Q. Adams, et~al., Science \textbf{390}(6777), 1029 (2025).
\newblock \doi{10.1126/science.adp6474}.
\newblock \urlprefix\url{https://doi.org/10.1126/science.adp6474}

\bibitem{CUORE:1TY}
D.Q. Adams, et~al., Nature \textbf{604}(7904), 53 (2022).
\newblock \doi{10.1038/s41586-022-04497-4}.
\newblock \urlprefix\url{https://doi.org/10.1038/s41586-022-04497-4}

\bibitem{Vetter:2023fas}
K.J. Vetter, et~al., Eur. Phys. J. C \textbf{84}(3), 243 (2024).
\newblock \doi{10.1140/epjc/s10052-024-12595-y}.
\newblock \urlprefix\url{https://doi.org/10.1140/epjc/s10052-024-12595-y}

\bibitem{adams2025endtoenddataanalysismethods}
D.Q. Adams, et~al.
\newblock End-to-end data analysis methods for the {CUORE} experiment (2025).
\newblock \urlprefix\url{https://arxiv.org/abs/2510.25720}

\bibitem{OptimumFilter}
E.~Gatti, P.F. Manfredi, La Rivista del Nuovo Cimento (1978-1999) \textbf{9}(1), 1 (1986).
\newblock \doi{10.1007/BF02822156}.
\newblock \urlprefix\url{https://doi.org/10.1007/BF02822156}

\bibitem{CUORE:analysisTech}
C.~Alduino, et~al., Phys. Rev. C \textbf{93}(4), 045503 (2016).
\newblock \doi{10.1103/PhysRevC.93.045503}.
\newblock \urlprefix\url{https://doi.org/10.1103/PhysRevC.93.045503}

\bibitem{Yocum:2022hum}
J.~Yocum, D.~Mayer, J.L. Ouellet, L.~Winslow, JINST \textbf{17}(07), P07004 (2022).
\newblock \doi{10.1088/1748-0221/17/07/P07004}.
\newblock \urlprefix\url{https://doi.org/10.1088/1748-0221/17/07/P07004}

\bibitem{pymoo}
J.~{Blank}, K.~{Deb}, IEEE Access \textbf{8}, 89497 (2020).
\newblock \doi{doi: 10.1109/ACCESS.2020.2990567}.
\newblock \urlprefix\url{https://doi.org/10.1109/ACCESS.2020.2990567}

\bibitem{Ramachandran2011}
P.~Ramachandran, G.~Varoquaux, CiSE \textbf{13}(2), 40–51 (2011).
\newblock \doi{10.1109/mcse.2011.35}.
\newblock \urlprefix\url{http://dx.doi.org/10.1109/MCSE.2011.35}

\bibitem{Geant4}
{Geant4 Collaboration}, IEEE Transactions on Nuclear Science \textbf{53}(1), 270 (2006).
\newblock \doi{10.1109/TNS.2006.869826}.
\newblock \urlprefix\url{https://doi.org/10.1109/TNS.2006.869826}

\bibitem{Geant4_2}
{Geant4 Collaboration}, NIMA \textbf{835}, 186 (2016).
\newblock \doi{10.1016/j.nima.2016.06.125}.
\newblock \urlprefix\url{https://doi.org/10.1016/j.nima.2016.06.125}

\bibitem{corner}
D.~Foreman-Mackey, JOSS \textbf{1}(2), 24 (2016).
\newblock \doi{10.21105/joss.00024}.
\newblock \urlprefix\url{https://doi.org/10.21105/joss.00024}

\bibitem{Bellini:2010iw}
F.~Bellini, et~al., JINST \textbf{5}, P12005 (2010).
\newblock \doi{10.1088/1748-0221/5/12/P12005}.
\newblock \urlprefix\url{https://doi.org/10.1088/1748-0221/5/12/P12005}

\bibitem{Andreino_Simonelli_2016}
A.~Simonelli, J.~Belﬁ, N.~Beverini, G.~Carelli, A.D. Virgilio, E.~Maccioni, G.D. Luca, G.~Saccorotti, Annals of Geophysics \textbf{59} (2016).
\newblock \doi{10.4401/ag-6970}.
\newblock \urlprefix\url{http://dx.doi.org/10.4401/ag-6970}

\bibitem{Mayer:2024fvd}
D.W. Mayer, {Advanced Reconstruction Techniques for CUORE: Searching Beyond the Standard Model with Cryogenic Calorimeters}.
\newblock Ph.D. thesis, MIT (2024).
\newblock \urlprefix\url{https://dspace.mit.edu/handle/1721.1/157595}

\bibitem{PhysRevD.100.062002}
N.Y. Agafonova, et~al., Phys. Rev. D \textbf{100}, 062002 (2019).
\newblock \doi{10.1103/PhysRevD.100.062002}.
\newblock \urlprefix\url{https://link.aps.org/doi/10.1103/PhysRevD.100.062002}

\bibitem{Woodley:2024eln}
W.~Woodley, A.~Fedynitch, M.C. Piro, Phys. Rev. D \textbf{110}(6), 063006 (2024).
\newblock \doi{10.1103/PhysRevD.110.063006}.
\newblock \urlprefix\url{https://doi.org/10.1103/PhysRevD.110.063006}

\bibitem{Borexino:2018pev}
M.~Agostini, et~al., JCAP \textbf{02}, 046 (2019).
\newblock \doi{10.1088/1475-7516/2019/02/046}.
\newblock \urlprefix\url{https://doi.org/10.1088/1475-7516/2019/02/046}

\bibitem{Bellini_2011}
G.~Bellini, et~al., JINST \textbf{6}(05), P05005–P05005 (2011).
\newblock \doi{10.1088/1748-0221/6/05/p05005}.
\newblock \urlprefix\url{http://dx.doi.org/10.1088/1748-0221/6/05/P05005}

\bibitem{Moore:2025eil}
M.~Moore, et~al., JINST \textbf{20}(08), P08020 (2025).
\newblock \doi{10.1088/1748-0221/20/08/P08020}.
\newblock \urlprefix\url{https://doi.org/10.1088/1748-0221/20/08/P08020}

\end{thebibliography}

\end{document}